\documentclass[final,5p,times,twocolumn]{elsarticle}

\usepackage{ecrc}

\usepackage{amsmath}
\usepackage{graphicx}
\usepackage{amssymb}
\usepackage{color,soul}
\usepackage[T1]{fontenc}
\usepackage{bm}
\usepackage{ae,aecompl}
\volume{}

\firstpage{1}

\journalname{}

\runauth{}

\jid{}

\jnltitlelogo{}

\usepackage{amssymb}
\begin{document}

\begin{frontmatter}



\dochead{}

\title{Transient dynamics of pulse-driven memristors in the presence of a stable fixed point}


\author[1]{Valeriy A. Slipko}
\ead{vslipko@uni.opole.pl}

\author[2]{Yuriy V. Pershin\corref{cor1}}
\ead{pershin@physics.sc.edu}

\cortext[cor1]{Corresponding author}
\address[1]{Institute of Physics, Opole University, Opole 45-052, Poland}
\address[2]{Department of Physics and Astronomy, University of South Carolina, Columbia, South Carolina 29208, USA}

\begin{abstract}
Some memristors are quite interesting from the point of view of dynamical systems.
When driven by narrow pulses of alternating polarities, their dynamics has a stable fixed point, which may be useful for future applications.
We study the transient dynamics of two types of memristors characterized by a stable fixed point using a time-averaged evolution equation.
Time-averaged trajectories of the Biolek window function memristor and resistor-threshold type memristor circuit (an effective memristor) are determined analytically, and the times of relaxation to the stable fixed point are found.
Our analytical results are in perfect agreement with the results of numerical simulations.
\end{abstract}


\end{frontmatter}

\section{Introduction} \label{sec:1}
Memristive devices~\cite{chua76a}, systems, and technologies are now common terms in the condensed matter and engineering literature.
During the past decade, memristive technologies have shown a bifurcating trail of early concept developments in the areas
of memristive neural networks~\cite{pershin09c,thomas2013memristor,prezioso2015training}, Boolean logic~\cite{borghetti10a,Sun11a,pershin19b}, and network computing~\cite{pershin11d,pershin13a}, which are various kinds of computing with memory~\cite{diventra13a}.
Together with memcapacitors and meminductors~\cite{diventra09a}, memristors offer a wide range of circuit functionalities that are not accessible with the traditional circuit components.
To develop reliable memristor circuit designs, it is indispensable to fully understand the response of individual (discrete) memristors and simple circuits made with them.

Recently, we have predicted stable fixed points in the dynamics of some pulse-driven memristors and their networks~\cite{pershin18b}.
In the subsequent study~\cite{slipko2018importance}, broad classes of memristor models characterized by a single stable fixed point, or
lack thereof, have been identified.
However, the questions of how long it will take for a memristor to approach a stable fixed point from an initial state, and what is the corresponding trajectory, have not been addressed so far.
Both of these are in the focus of the present investigation.

Before continuing, it is necessary to introduce the concept of an $n$th order
current-controlled memristive system (memristor), as defined by~\cite{chua76a}:
\begin{eqnarray}
V_M(t)&=&R_M\left(\boldsymbol{x},I \right)I(t), \label{eq1}\\
\dot{\boldsymbol{x}}&=&\boldsymbol{f}\left(\boldsymbol{x},I\right), \label{eq2}
\end{eqnarray}
where $V_M$ and $I$ are the voltage across and current through the memristor, respectively, $R_M\left( \boldsymbol{x}, I\right)$ is the memristance (memory resistance), $\boldsymbol{x}$ is a vector of $n$ internal state variables, and $\boldsymbol{f}\left(\boldsymbol{x}, I \right)$ is a vector function\footnote{Voltage-controlled memristive systems are defined similarly~\cite{chua76a}.}.
In this paper, we will consider only first-order memristor models, so that in what follows, $\boldsymbol{x}$ and $\boldsymbol{f}\left(\boldsymbol{x}, I \right)$ are scalars, $x$ and $f(x,I)$, respectively.

Our main interest is to understand the general behavior of
memristors  subjected to narrow pulses of current or voltage with alternating polarities.
For this purpose, we use a time-averaged version of Eq.~(\ref{eq2}) to describe the evolution of the internal state variable averaged over the pulse period $T$.
 Below, we consider two specific cases: a Biolek window function memristor subjected to current pulses and a resistor-threshold type memristor circuit subjected to voltage pulses.
In both cases, the evolution equation is solved exactly and characteristic relaxation times
describing the approach to the stable fixed point are found.
Our main results related to the Biolek window function memristor and resistor-threshold type memristor circuit are given by Eqs.~(\ref{eq:bio5}) and (\ref{eq:bio7}), and Eqs.~(\ref{eq:41}) and  (\ref{eq:43}), respectively.

\begin{figure}[b]
\centering{\includegraphics[width=65mm]{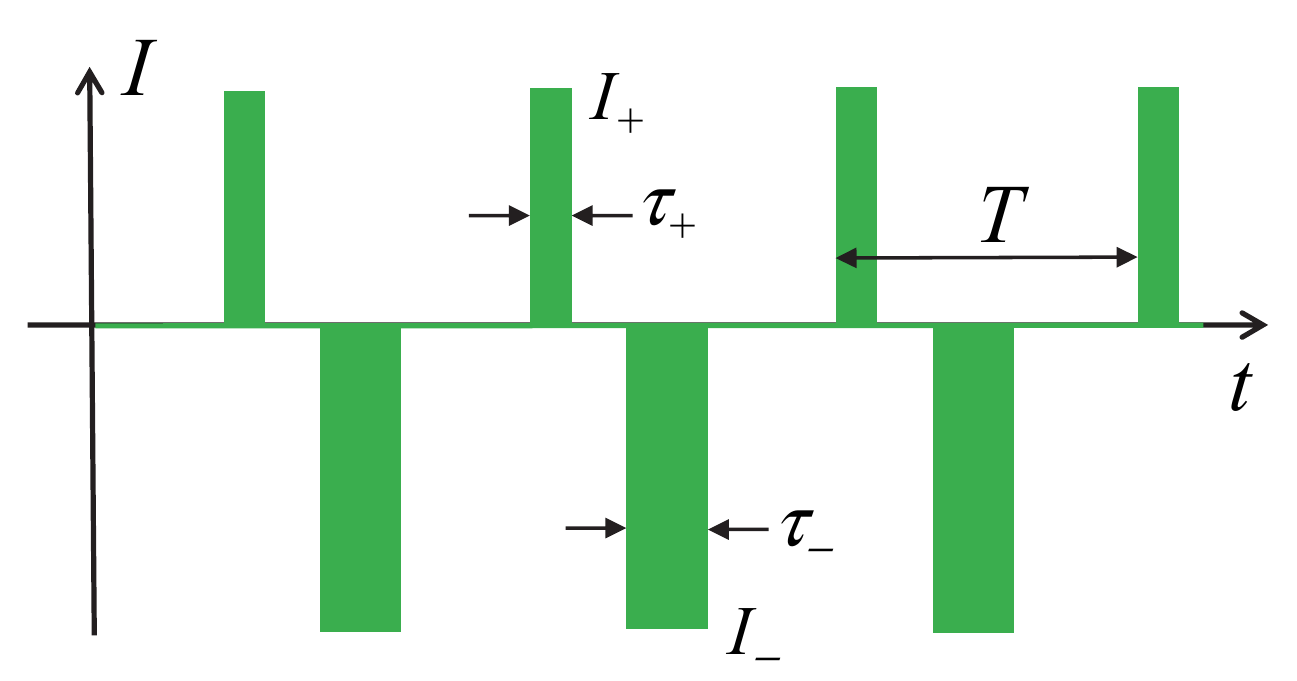}}
\caption{Pulse sequence schematics.
Here, $T$ is the period, $I_+$ and $I_-$ are the amplitudes of
the positive and negative pulses of current, respectively, $\tau_+$ and $\tau_-$ are their widths.}\label{fig:1}
\end{figure}

This paper is organized as follows.
Section \ref{sec:2} introduces a general framework for the study of time-averaged dynamics of driven memristors.
In Section \ref{sec:3}, we investigate the transient dynamics of the
Biolek window function memristor and resistor-threshold type memristor circuit.
We conclude in Section
\ref{sec:4} with a discussion.

\section{Description of the time-averaged dynamics} \label{sec:2}

For the convenience of the reader, in this section we introduce an equation describing the time-averaged dynamics of memristors driven by short alternating-polarity pulses of current~\cite{slipko2018importance}.
In the derivation below, the central assumption  is that the change in the internal state variable $x$ induced by each pulse is small.
Schematics of the sequence of pulses are shown in Fig.~\ref{fig:1}, which also defines the parameters of the sequence of pulses.

Generally speaking, the evolution of a pulse-driven memristor consists of a slow drift and fast oscillations.
As our primary interest is in understanding the behavior of the memristor on the whole, we integrate over the fast oscillations by introducing a time-averaged internal state variable $\bar{x}$:
\begin{equation}\label{eq:average}
  \bar{x}(t)=\frac{1}{T}\int\limits_{t}^{t+T} x(\tau)\textnormal{d}\tau.
\end{equation}
On the one hand, by differentiating Eq. (\ref{eq:average}), we get
\begin{equation}\label{eq:average1}
  \dot{\bar{x}}(t)=\frac{x(t+T)-x(t)}{T}.
\end{equation}
On the other hand, the integration of Eq. (\ref{eq2}) from $t$ to $t+T$ leads to
\begin{equation}\label{eq:average2}
 x(t+T)-x(t)= f(\bar{x},I_+)\tau_++f(\bar{x},I_-)\tau_-+f(\bar{x},0)\tau_0,
\end{equation}
where $\tau_0=T-\tau_+-\tau_-$.
In what follows, we set $f(\bar{x},0)=0$, which is typically satisfied in the
first-order non-volatile memristor models.

Combining Eqs.~(\ref{eq:average1}) and (\ref{eq:average2}), the equation for the time-averaged evolution can be written as
\begin{equation}\label{eq:average3}
  \dot{\bar{x}}(t)=\frac{1}{T}\left(f(\bar{x},I_+)\tau_++f(\bar{x},I_-)\tau_-\right).
\end{equation}
Eq.~(\ref{eq:average3}) forms the basis for our analysis.
We emphasize that by equating  the right-hand side of Eq.~(\ref{eq:average3}) to zero,
we obtain an equation for a fixed point.
Moreover, the fixed point is stable if, at its location, the first derivative of the right-hand side of Eq.~(\ref{eq:average3})
with respect to $\bar{x}$ is negative,
\begin{equation}\label{eq:average4}
\left.\frac{\partial f(\bar{x},I_+)}{\partial \bar{x}}\right|_{\bar{x}=x_a}\tau_++\left.\frac{\partial f(\bar{x},I_-)}{\partial \bar{x}}\right|_{\bar{x}=x_a}\tau_- <0,
\end{equation}
where $x_a$ is the location of the fixed point in question.

We note that the above derivation is also valid for the case of voltage-controlled memristors driven by voltage pulses, with a corresponding replacement of current by voltage.

\section{Results} \label{sec:3}

\subsection{Biolek window function memristor}\label{sec:31}

Here we consider the approach to a stable fixed point of
a current-controlled memristor subjected to narrow pulses of current with alternating polarities.
Based on our previous work~\cite{slipko2018importance}, we assume that the memristor
is described by a Biolek window function~\cite{Biolek2009-1} model, such that
$f(x,I)$ in Eq.~(\ref{eq2}) is of the form
\begin{equation}
  f(x,I)=h(I)g_B(x,I), \label{eq:bio1}
\end{equation}
where $h(I)$ is a function of the current, $g_B(x,I)$ is the Biolek window function~\cite{Biolek2009-1}, $h(I)\geq 0$ for $I>0$, $h(I)\leq 0$ for $I<0$, and $h(I)= 0$ for $I=0$.
The role of $g_B(x,I)$  is to restrict the change of the
internal state variable to the region between 0 and 1.
According to \cite{Biolek2009-1},
\begin{equation}\label{eq:bio2}
g_B(x,I)=1-(x-H(-I))^{2p},
\end{equation}
where $H(...)$ is the Heaviside step function and $p$ is a positive integer (in what follows, we take $p=1$).
Previously, we have demonstrated that there is always a single stable fixed point in the dynamics of such a memristor~\cite{slipko2018importance}.

\begin{figure*}[t]
(a) \centering{\includegraphics[width=70mm]{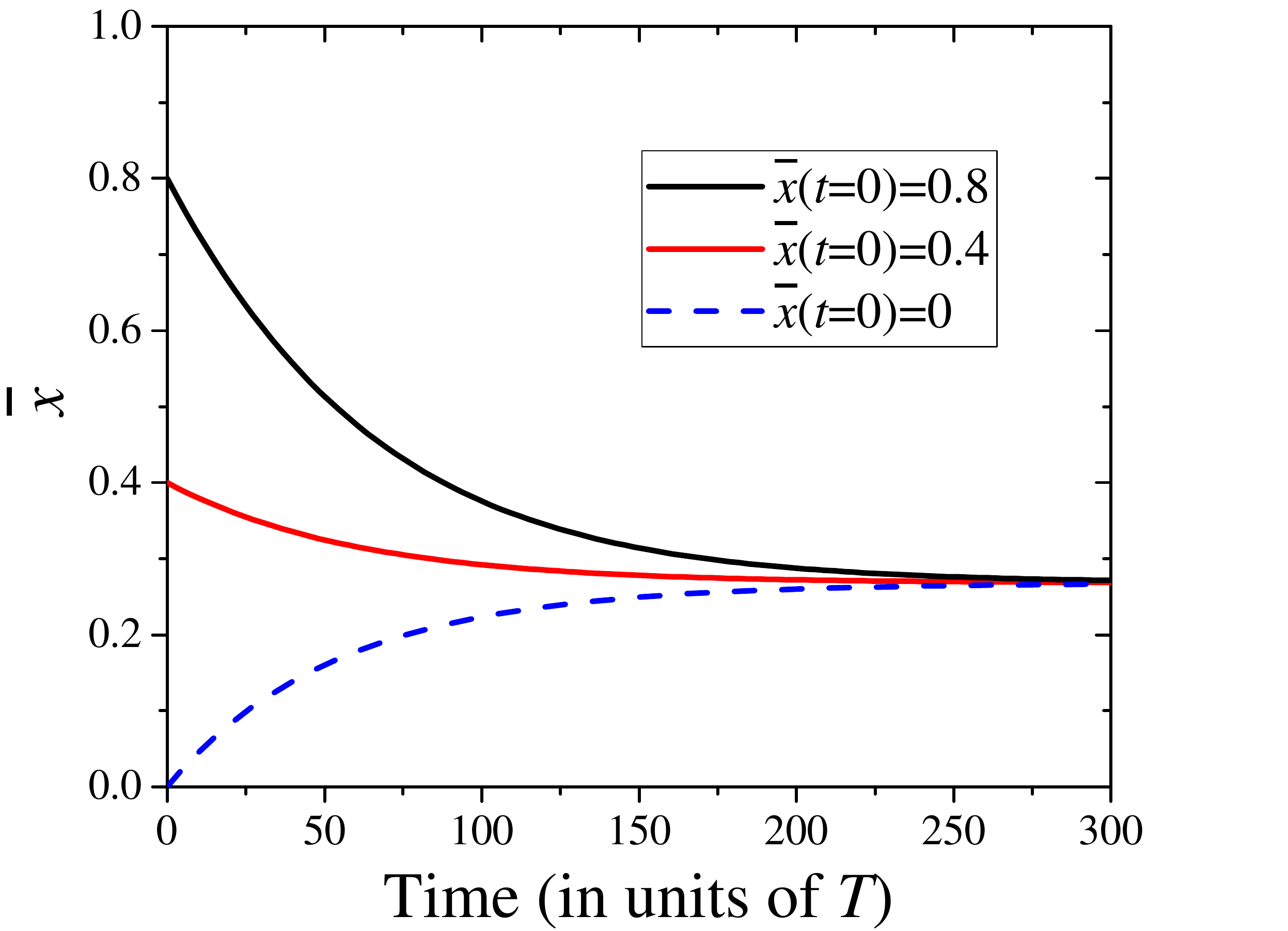}} \;\;\;\;
(b) \centering{\includegraphics[width=70mm]{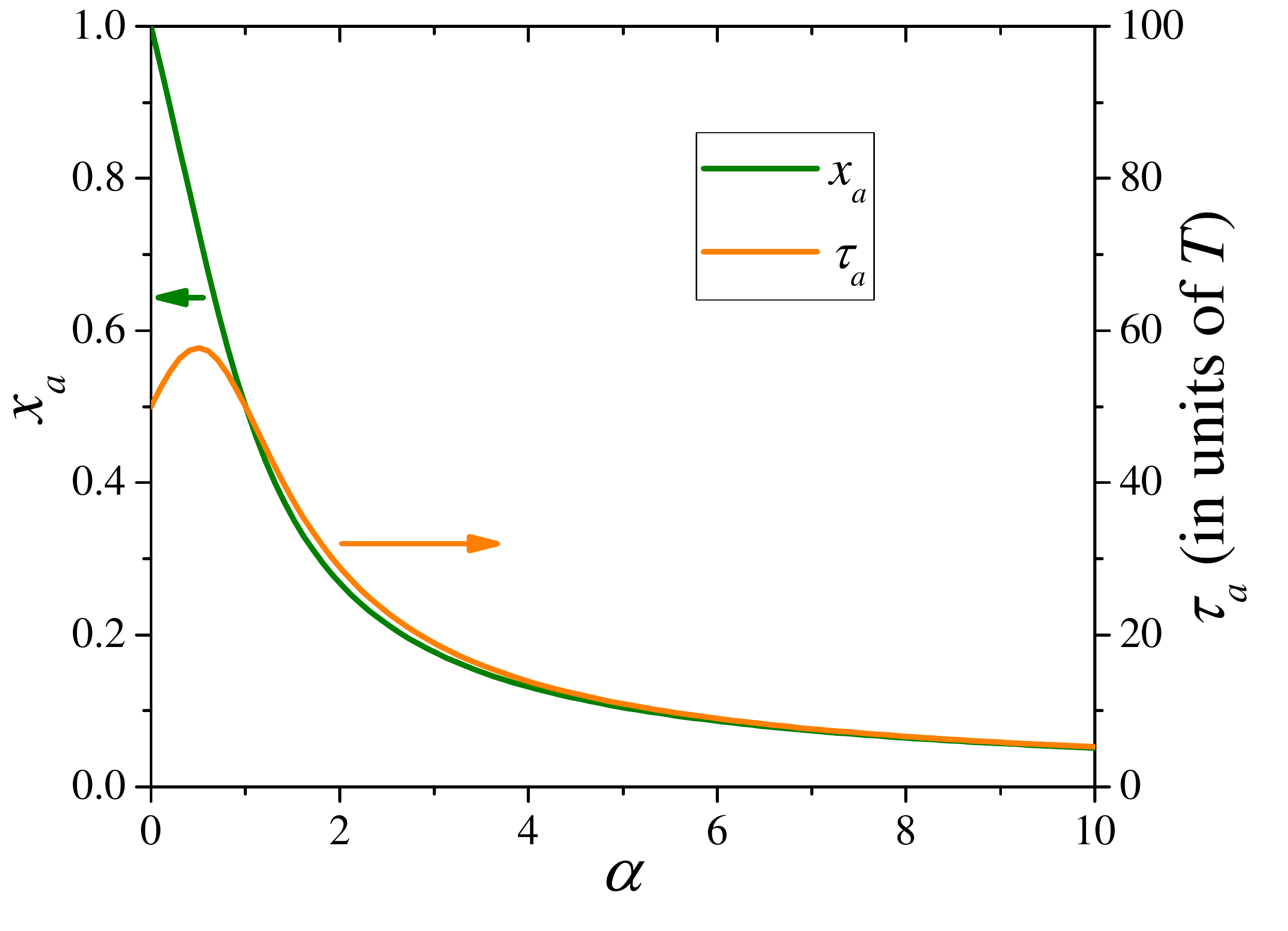}}
\caption{ Transient dynamics of the Biolek window function memristor.
(a) Time dependence of the time-averaged internal state variable for different initial conditions: $\alpha=2$; $h(I_+)\tau_+=0.01$.
(b) Position of the stable fixed point, $x_a$, and characteristic relaxation time, $\tau_a$, as functions of $\alpha$; $h(I_+)\tau_+=0.01$.
These plots were obtained using (a) Eq. (\ref{eq:bio5}), and (b) Eqs.~(\ref{eq:bio6}) and (\ref{eq:bio7}).}
\label{fig:2}
\end{figure*}

To fully define the memristor model, the memristance $R_M$ as a function of $x$ and $I$ needs to be specified (see Eq. (\ref{eq1})).
In this paper we suppose there is a linear relation between $R_M$ and $x$ (the same for both models),
\begin{equation}\label{eq:33}
  R_M(x)=R_{\textnormal{off}}+x\left( R_{\textnormal{on}}-R_{\textnormal{off}}\right),
\end{equation}
where $R_{\textnormal{off}}$ and $R_{\textnormal{on}}$ are the high- and low-resistance states of the memristor (the ``off'' and ``on'' memristor states, respectively), and $x$ is bound to the region between 0 and 1.

Using Eqs.~(\ref{eq:average3}) and (\ref{eq:bio1}), and assuming that $h(I_+)> 0$ and $h(I_-)< 0$, the time-averaged memristor dynamics
is described by the equation
\begin{equation} \label{eq:bio3}
\dot{\bar{x}}(t)=\frac{h(I_+)\tau_+}{T}\left[ (\alpha-1)\bar{x}^2-2\alpha\bar{x}+1 \right],
\end{equation}
where $\alpha=|h(I_-)\tau_-|/(h(I_+)\tau_+)$ is a positive constant.
When $\alpha = 1$, the solution of Eq. (\ref{eq:bio3}) can be written in the form
\begin{equation} \label{eq:bio4}
\bar{x}(t)=\frac{1}{2}+\left( x_0-\frac{1}{2}\right)\exp\left(-\frac{2h(I_+)\tau_+}{T}t\right).
\end{equation}
Here, $\bar{x}(t=0)=x_0$ is the initial condition.
According to Eq.~(\ref{eq:bio4}), the time-averaged internal state variable approaches exponentially the stable fixed point located at $x_a=1/2$.

When $\alpha\neq 1$, the
solution of Eq.~(\ref{eq:bio3}) can be presented as
\begin{equation} \label{eq:bio5}
\bar{x}(t)=
\frac{
(\alpha x_0-1)
\textnormal{tanh}
\left[\frac{h(I_+)\tau_+D[\alpha]}{T}t\right]
-D[\alpha]x_0
}
{\left[(\alpha-1)x_0-\alpha\right]
\textnormal{tanh}
\left[\frac{h(I_+)\tau_+D[\alpha]}{T}t\right]
-D[\alpha]
},
\end{equation}
where we write $D[\alpha]=\sqrt{1-\alpha+\alpha^2}$ for brevity.
Eq.~(\ref{eq:bio5}) describes the evolution of
$\bar{x}(t)$ towards the stable fixed point at
\begin{equation}\label{eq:bio6}
x_a=\frac{\alpha-\sqrt{1-\alpha+\alpha^2}}{\alpha-1}.
\end{equation}
It follows from Eq.~(\ref{eq:bio5}) that the characteristic relaxation time is
\begin{equation}\label{eq:bio7}
\tau_a= \frac{T}{2h(I_+)\tau_+D[\alpha]},
\end{equation}
where the factor $1/2$ is used to match the relaxation time in Eq.~(\ref{eq:bio4}) representing the solution at $\alpha=1$.

Fig.~\ref{fig:2}(a) shows an example of the stable fixed point dynamics of the Biolek window function memristor.
This plot was obtained for few selected initial conditions of memristor.
The location of the fixed point (Eq.~(\ref{eq:bio6})) and characteristic relaxation time (Eq.~(\ref{eq:bio7})) as functions of $\alpha$ are presented in
Fig.~\ref{fig:2}(b).
We emphasize that the relaxation time $\tau_a$ is independent of the initial state of the memristor.

It is interesting to note that the time-averaged dynamics of the Biolek window function memristor possesses a symmetry with respect to the transformation $\alpha \rightarrow 1/\alpha$, $\bar{x}\rightarrow 1-\bar{x}$, $t\rightarrow \alpha t$.
This follows from Eq.~(\ref{eq:bio3}) or directly from the solution (\ref{eq:bio5}).
Therefore, in principle, the study of such a memristor could be limited to the finite region $0<\alpha\leq 1$ of the parameter $\alpha$, and the above symmetry could be used to extend the solution to the region $\alpha>1$.

\subsection{Resistor-threshold type memristor circuit}\label{sec:32}

Next we investigate the approach to a stable fixed
point of a voltage-controlled memristor subjected to narrow voltage pulses with alternating polarities.
As an effective memristor, we consider a resistor-threshold type memristor circuit.
According to \cite{pershin18b}, there can be a fixed
point attractor in its dynamics.

\begin{figure*}[t]
(a) \centering{\includegraphics[width=70mm]{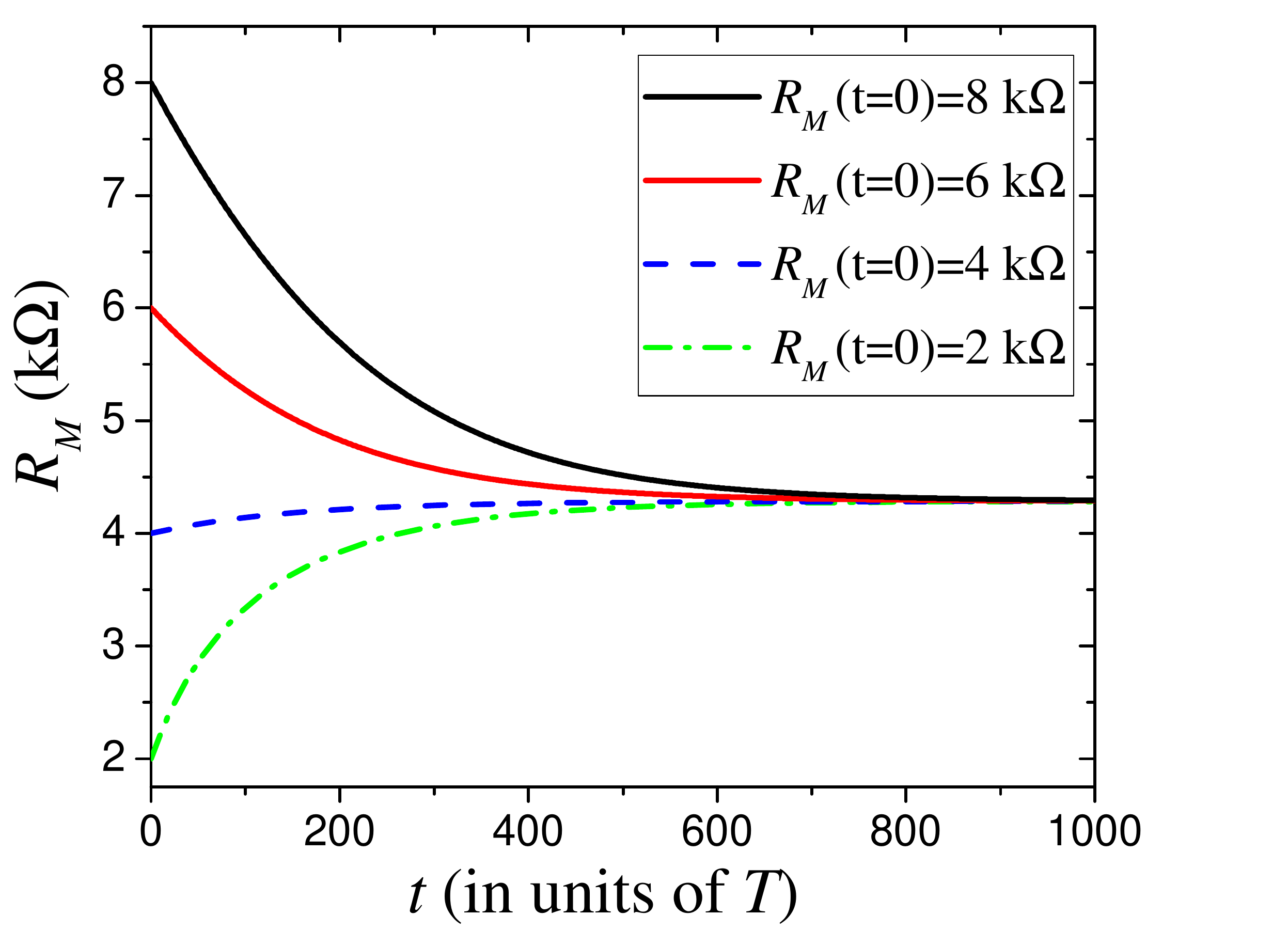}} \;\;\;\;
(b) \centering{\includegraphics[width=70mm]{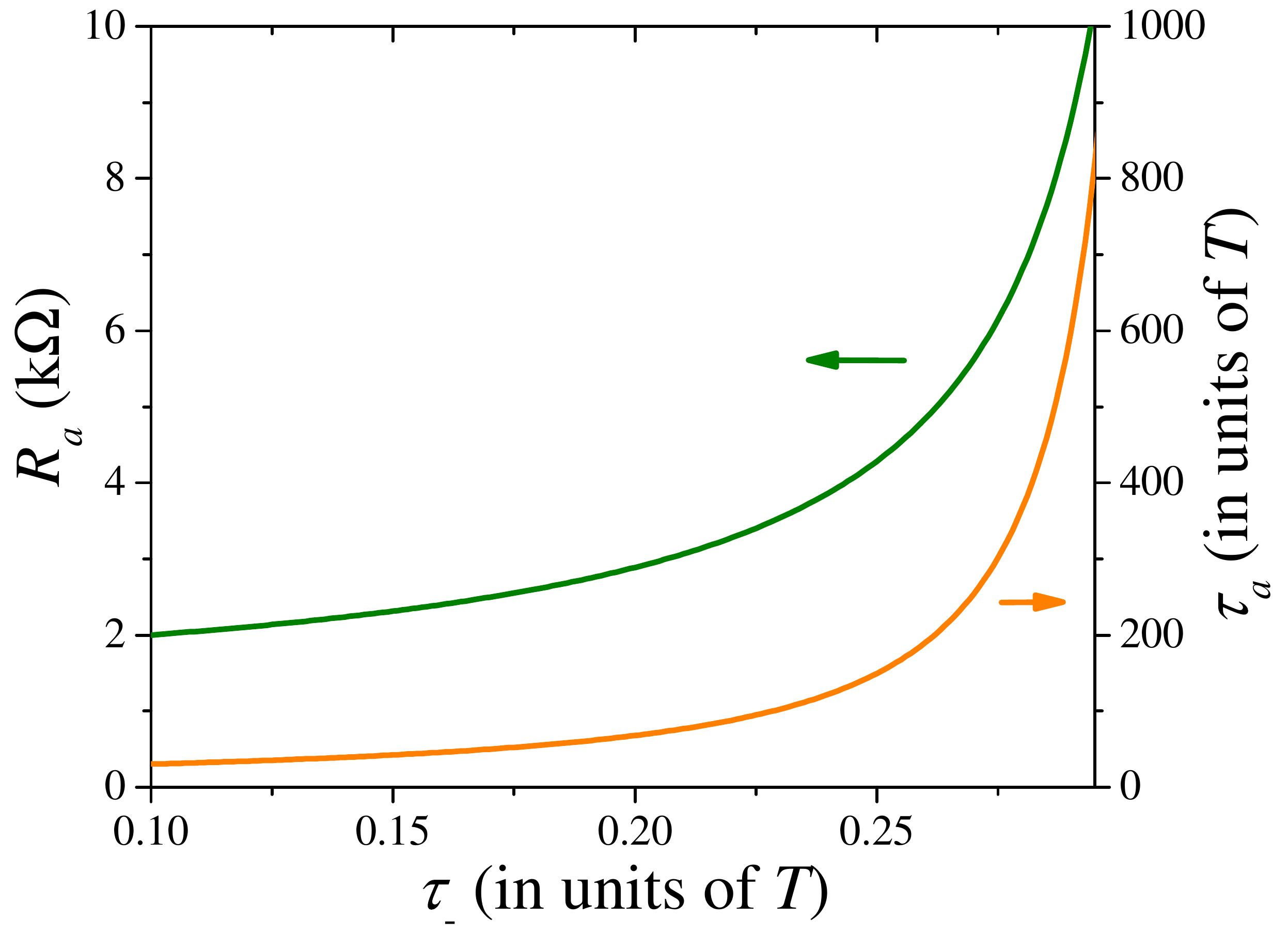}}
\caption{Transient dynamics of the resistor-threshold type memristor circuit.
(a) Memristance as a function of time for different initial conditions.
(b) Location of the stable fixed point $R_a$ and the characteristic relaxation time $\tau_a$ as functions of $\tau_-$.
These plots were obtained using (a) Eq.~(\ref{eq:43}), and (b) Eqs.~(\ref{eq:37}) and (\ref{eq:41}); $V_+=-V_-=2.2$~V; $\tau_+=0.4T$; $\tau_-=0.25T$ in (a); $V_{\textnormal{on}}=1$~V; $V_{\textnormal{off}}=-0.7$~V; $R=R_{\textnormal{on}}=2$~k$\Omega$; $R_{\textnormal{off}}=10$~k$\Omega$; $\beta T=0.05$~V$^{-1}$.}
\label{fig:3}
\end{figure*}

Our study is based on a simple threshold-type memristor model introduced in \cite{pershin09b}:
\begin{equation}\label{eq:30}
  \dot{x}=\left\{
                \begin{array}{lll}
                  \beta\left( V_M-V_{\textnormal{on}} \right) &, &  0<V_{\textnormal{on}}<V_M\\
                  0 &, &  V_{\textnormal{off}}<V_M<V_{\textnormal{on}} \quad.\\
                  \beta \left(V_M - V_{\textnormal{off}} \right)&, &    V_M<V_{\textnormal{off}}<0
                \end{array}
              \right.
\end{equation}
Here, $\beta$ is a positive constant, while $V_{\textnormal{on}}$ and $V_{\textnormal{off}}$ are the positive and negative threshold voltages, respectively.  It is assumed that $R_M$ and $x$ are connected via Eq.~(\ref{eq:33}).

Next, consider the threshold-type memristor connected in series with a resistor $R$.
Then, Eq.~(\ref{eq:30}) can be rewritten as
\begin{equation}\label{eq:31}
  \dot{x}=\left\{
                \begin{array}{lll}
                  \beta\left( \frac{R_M(x)}{R+R_M(x)}V(t)-V_{\textnormal{on}} \right)  &, & 0<V_{\textnormal{on}}<V_M\\
                  0 &, & V_{\textnormal{off}}<V_M<V_{\textnormal{on}}, \\
                  \beta \left(\frac{R_M(x)}{R+R_M(x)}V(t) - V_{\textnormal{off}} \right) &, &  V_M<V_{\textnormal{off}}<0
                \end{array}
              \right.
\end{equation}
where $V(t)$ is the externally applied voltage.
Assuming that the pulse amplitudes are high enough so that the voltage
across the memristor is always above its threshold, the time-averaged evolution equation has the form
\begin{eqnarray}\label{eq:32}
\dot{\bar{x}}(t)=\frac{\beta}{T}\left[\left( \frac{R_M(\bar{x})}{R+R_M(\bar{x})}V_+-V_{\textnormal{on}} \right)\tau_+ \right.
\nonumber \\
\qquad \qquad  \left.
+  \left( \frac{R_M(\bar{x})}{R+R_M(\bar{x})}V_--V_{\textnormal{off}} \right)\tau_- \right].
\end{eqnarray}
Then Eq.~(\ref{eq:32}) may be rewritten as
\begin{equation}\label{eq:34}
\dot{\bar{x}}(t)=\frac{\beta}{T}\left[ \frac{R_M(\bar{x})}{R+R_M(\bar{x})}\kappa -p\right],
\end{equation}
where $\kappa=V_+\tau_++V_-\tau_-$ and $p=V_{\textnormal{on}}\tau_++V_{\textnormal{off}}\tau_-$.

Taking into account Eq.~(\ref{eq:33}), Eq.~(\ref{eq:34}) can be presented as
\begin{equation}\label{eq:35}
\frac{\textnormal{d}R_M}{\textnormal{d} t}=-\frac{\beta \left( R_{\textnormal{off}}-R_{\textnormal{on}} \right)}{T}\left[ \frac{R_M}{R+R_M}\kappa -p \right],
\end{equation}
where, for the sake of simplicity, we use $R_M\equiv R_M(\bar{x})$.
Consider the fixed points
of Eq.~(\ref{eq:35}).
It is evident that these correspond to
\begin{equation}\label{eq:36}
   \frac{R_M}{R+R_M}\kappa -p=0,
\end{equation}
which leads to
\begin{equation}\label{eq:37}
   R_a=R\frac{p}{\kappa - p}.
\end{equation}
Note that for some choices of parameters for the pulse sequence, the fixed point $R_a$ determined by Eq.~(\ref{eq:37}) may be located outside the interval $[R_{\textnormal{on}},R_{\textnormal{off}}]$ of allowed memristances.
In such a case, the memristance  changes monotonically towards one of the limiting values ($R_{\textnormal{on}}$ or $R_{\textnormal{off}}$), depending on the sign of the right-hand side of Eq.~(\ref{eq:35}).

Let us first consider the dynamics close to the fixed point, supposing that $R_{\textnormal{on}}<R_a<R_{\textnormal{off}}$.
Substituting $R_M=R_{a}+\delta R_M$ into Eq.~(\ref{eq:35}) and
expanding up to linear terms we get
\begin{equation}\label{eq:38}
  \frac{\textnormal{d}}{\textnormal{d}t}\delta R_M=-\frac{\beta (R_{\textnormal{off}}-R_{\textnormal{on}})}{T}\frac{\kappa R}{\left( R+R_a \right)^2}\delta R_M.
\end{equation}
The solution is
\begin{equation}\label{eq:39}
  \delta R_M(t)=\delta R_M(0)\cdot \exp\left\{ -\frac{\beta (R_{\textnormal{off}}-R_{\textnormal{on}})}{T}\frac{\kappa R}{\left( R+R_a \right)^2} t  \right\} .
\end{equation}
It follows from the above equation that if
\begin{equation}\label{eq:40}
 % \gamma=
  \frac{\beta (R_{\textnormal{off}}-R_{\textnormal{on}})}{T}\frac{\kappa R}{\left( R+R_a \right)^2}>0,
\end{equation}
then the fixed point is stable. An obviously sufficient condition is that $\kappa>0$.

Using Eqs.~(\ref{eq:37}) and (\ref{eq:39}) we
obtain the characteristic  relaxation time for the resistor-threshold type memristor circuit:
\begin{equation}\label{eq:41}
  \tau_a=\frac{T\kappa R}{\beta (R_{\textnormal{off}}-R_{\textnormal{on}}) \left(\kappa-p \right)^2}.
\end{equation}

It follows from Eq.~(\ref{eq:39}) that in the case of a stable fixed point $(\kappa>0)$, all initial memristances from at least some vicinity of $R_a$ approach asymptotically and monotonically the location of the fixed point in the limit as $t \rightarrow +\infty$.
In fact, this is true for any initial condition of Eq.~(\ref{eq:35}) (based on the analysis of the sign of its right-hand side).
Thus any initial condition belongs to the basin of the single fixed point specified by Eq.~(\ref{eq:37}).

Lastly, we will derive the exact solution of Eq.~(\ref{eq:35}).
This equation can be rewritten as
\begin{equation}\label{eq:42}
  \frac{R_M+R}{R_M-R_a}\textnormal{d}R_M=-\frac{\beta (R_{\textnormal{off}}-R_{\textnormal{on}})}{T}(\kappa-p)\textnormal{d}t.
\end{equation}
Upon integrating we obtain
\begin{eqnarray}
  R_M-R_M(0)+(R+R_a)\ln\frac{R_M-R_a}{R_M(0)-R_a}= \nonumber \\
  -\frac{\beta (R_{\textnormal{off}}-R_{\textnormal{on}})}{T}(\kappa-p)t.
\label{eq:43}
\end{eqnarray}
Eq.~(\ref{eq:43}) implicitly defines
 $R_M=R_M(t)$.

Fig.~\ref{fig:3} illustrates the analytical results obtained in this subsection.
It shows, in particular, (a) the transient dynamics of the circuit, and (b) the location  of the attractor point, and the relaxation time, as functions of the pulse parameter $\tau_-$ (all other parameters being kept fixed).

\section{Discussion and conclusion} \label{sec:4}

\begin{figure}[t]
(a)\centering{\includegraphics[width=70mm]{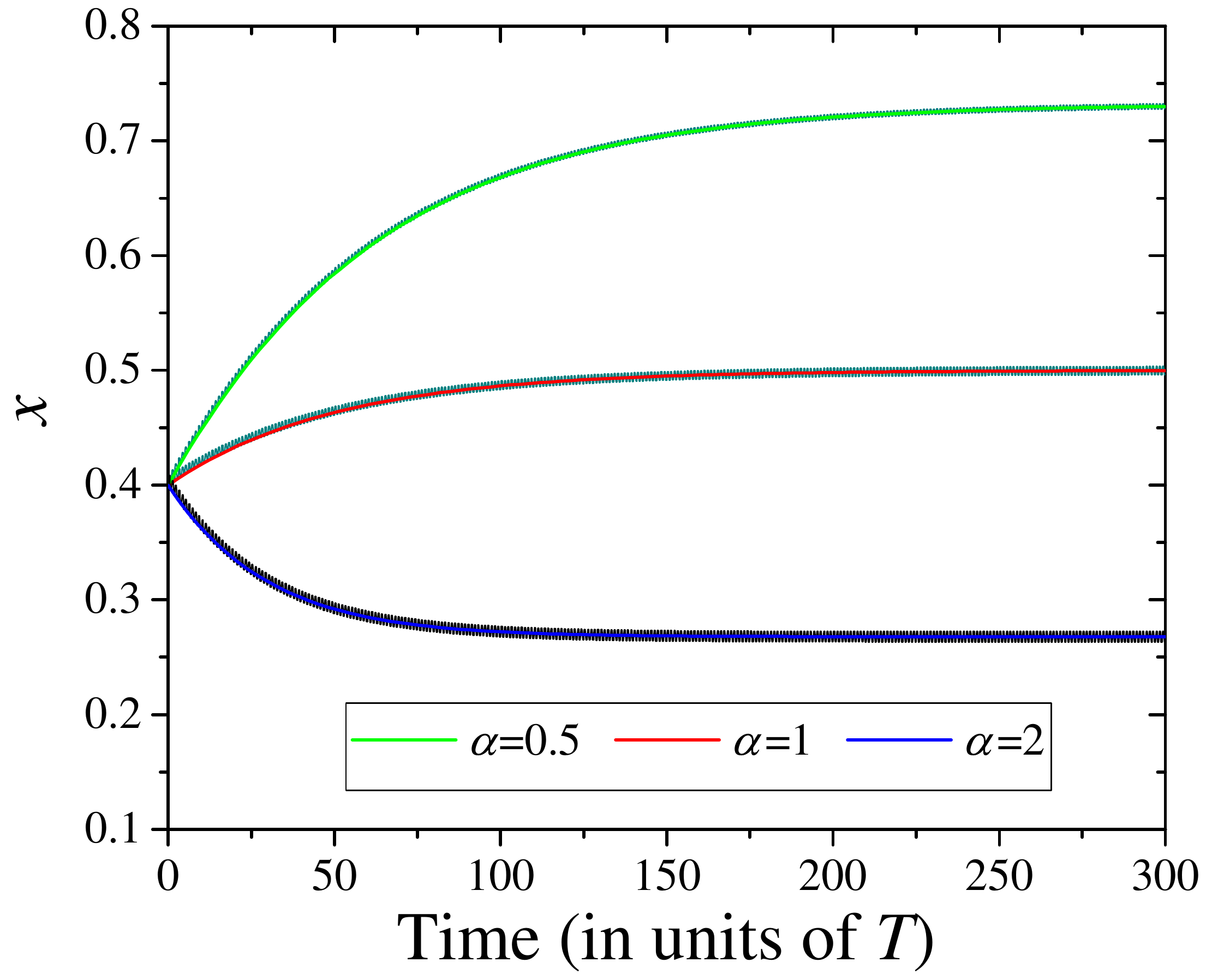}} \\
(b)\centering{\includegraphics[width=70mm]{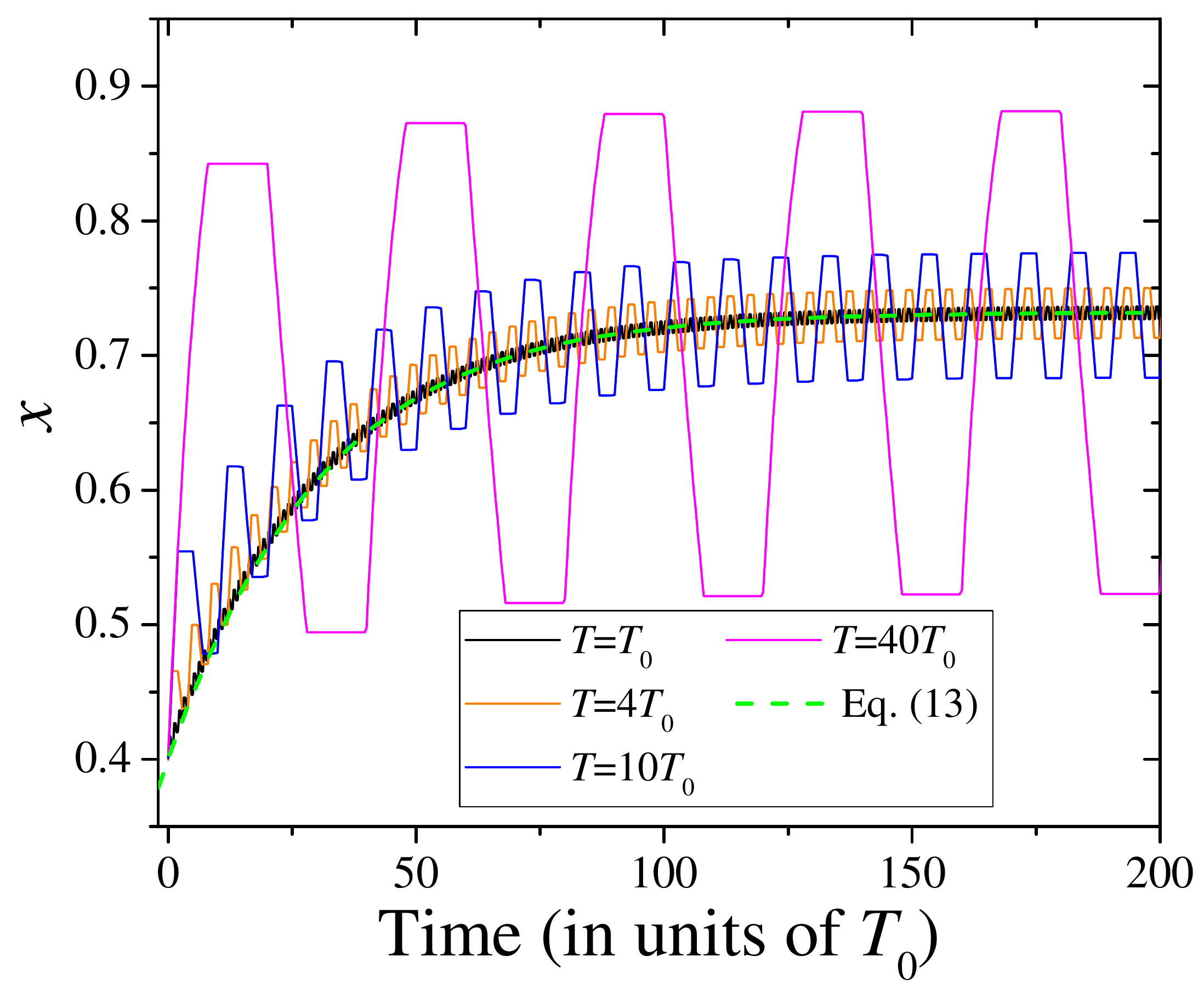}}
\caption{Comparison of the exact dynamics and time-averaged solution in the Biolek window function memristor.
(a) Trajectories for different values of $h(I_-)$;
$\tau_+=\tau_-=0.2T$; $h(I_+)\tau_+=0.01$.
The smooth curves representing $\bar{x}$ were obtained using Eq.~(\ref{eq:bio5}).
The `noisy' curves are the numerical solution.
(b) Trajectories for different values of the pulse period $T$; $\tau_+=\tau_-=0.2T$; $\alpha=0.5$.
The period $T_0$ is defined via $|h(I_-)0.2T_0|=0.01$.
}\label{fig:4}
\end{figure}

In this paper, the time-averaged evolution equation describing the dynamics of pulse-driven memristors was solved analytically in two model cases.
We have shown that
in the Biolek window function memristor, the trajectories converge to the stable fixed point as a hyperbolic tangent function of time.
Moreover, the relaxation is exponential in the special case of symmetric pulses $h(I_+)\tau_+=|h(I_-)\tau_-|$.
In the case of the resistor-threshold type memristor circuit, we have derived
an implicit solution of the evolution equation.
Additionally, it has been shown that close to the stable fixed point, the relaxation is exponential.
The relaxation times have been identified for both types of memristors considered in this paper.

Our theory of the time-averaged evolution is based on the assumption of small changes in the memristor state induced by each pulse.
 To evaluate how close the  time-averaged trajectories are
to the exact ones, we have numerically simulated the dynamics of pulse-driven
memristors (using Eq.~(\ref{eq2})), and compared the exact and time-averaged trajectories.
An excellent agreement between the time-averaged theory and exact evolution has been found.
For instance, Fig.~\ref{fig:4}(a) shows the time-averaged and exact trajectories of the Biolek window function memristor for several representative driving conditions.
According to Fig.~\ref{fig:4}(a), the time-averaged and exact trajectories are almost indistinguishable from each other.

Moreover, we stress that the time-averaged equations may provide a very good description for the dynamics on the whole even outside the limits of the central assumption of the time-averaged approximation (see Section  \ref{sec:2}).
To show this, the exact trajectories were found numerically for several values of the pulse sequence period $T$ and compared to the predictions of the time-averaged approach.
Fig.~\ref{fig:4}(b) shows that even in the case of large oscillations, the time-averaged behavior is reasonably well described by $\bar{x}$.
An interesting observation is that the scaling of the pulse period does not change the relaxation time -- which can be inferred from the general form of the time-averaged evolution equation (\ref{eq:average3}).

Our findings can be used to describe the dynamics of physical memristors in the appropriate experimental conditions.
Moreover, this work is relevant to the recently studied `fading memory effect'~\cite{Ascoli16a,Menzel17a}, although the driving conditions considered in the present paper are different.
 The use of narrow fixed-amplitude pulses (instead of sinusoidal or triangular waveforms) is highly beneficial from the theoretical standpoint since the calculations are significantly simplified.

\bibliographystyle{elsarticle-num}
\bibliography{memcapacitor}

\end{document}